\documentstyle[./aabib,./psfig,./aasms4]{article} 
\def\Chang-Hwan#1{{\bf[#1 -- Chang-Hwan]}}

\newcommand{\phik}{\mbox{$\phi_{\rm k}$}}
\newcommand{\thetak}{\mbox{$\theta_{\rm k}$}}
\newcommand{\phio}{\mbox{$\phi_{\rm obs}$}}
\newcommand{\thetao}{\mbox{$\theta_{\rm obs}$}}
\newcommand{\phij}{\mbox{$\phi_{\rm jet}$}}
\newcommand{\thetaj}{\mbox{$\theta_{\rm jet}$}}

\def\arraystretch{1.1}
\def\be {\begin{eqnarray}}
\def\ee {\end{eqnarray}}

\def\ba {\begin{array}}
\def\ea {\end{array}}

\def\ben{\begin{enumerate}}
\def\een{\end{enumerate}}
\def\bi{\begin{itemize}}
\def\ei{\end{itemize}}
\def\I{{\cal I}}
\long\def\beginomit#1\endomit{}

\def\nn{\nonumber\\}

\newcommand{\msun}{\mbox{${\rm M}_\odot$}}

\newcommand{\rsun}{\mbox{${\rm R}_\odot$}}
\newcommand{\kms}{\mbox{${\rm km~s}^{-1}$}}
\newcommand{\kmsp}{\mbox{${\rm km~s.}^{-1}$}\,}

\def\unit#1{{\mbox{[{\rm #1}]}}}
\def\apgt{\ {\raise-.5ex\hbox{$\buildrel>\over\sim$}}\ }
\def\aplt{\ {\raise-.5ex\hbox{$\buildrel<\over\sim$}}\ }

\begin{document}

%\baselineskip 1.5 \baselineskip

\title{Can precessing jets explain the light curves of Gamma-ray Bursts?}

\author{
	Simon F.\ Portegies Zwart\altaffilmark{1} 
	and
	Chang-Hwan Lee\altaffilmark{2} 
	and
	Hyun Kyu Lee\altaffilmark{2, 3}
} 

\slugcomment{Simon Portegies Zwart is a Japan Society for the
	     Promotion of Science Fellow} 
\authoremail{spz@grape.c.u-tokyo.ac.jp}

\lefthead{Portegies Zwart et al.}
\righthead{Temporal structure of Gamma-ray bursts}

\altaffiltext{1}{
	Department of Information Science and Graphics, 
		College of Arts and Science, 
		University of Tokyo, 3-8-1 Komaba,
		Meguro-ku, Tokyo 153, Japan
}
\altaffiltext{3}{
	Department of Physics and Astronomy,
		State University of New York at Stony Brook,
                Stony Brook, NY11794, USA 
}
\altaffiltext{4}{
	Department of Physics, Hanyang University, Seoul 133-791, Korea
}

\keywords{05 ---
	accretion discs -- 
	black hole physics --
	binaries: close ---
	stars: neutron --
	gamma-rays: bursts --
	gamma-rays: theory --
%	instabilities ---
%	blue stragglers ---
%	open clusters and associations: general ---
%	stars: evolution ---
%	stars: kinematics ---
%	stars: mass-loss 
%	stars: Wolf-Rayet ---
%	globular clusters: general ---
%	galaxies: star clusters ---
%	galaxies: kinematics and dynamics
	}  

\begin{abstract}

We present a phenomenological model to explain the light curves of
gamma-ray bursts. In the model a black hole is orbited by a precessing
accretion disc which is fed by a neutron star. Gamma-rays are produced
in a highly collimated beam via the Blandford-Znajek mechanism.  The
gamma-ray beam sweeps through space due to the precession of the
slaved accretion disc.  The light curve expected from such a
precessing luminosity cone can explain the complex temporal behavior
of observed bright gamma-ray bursts.

\end{abstract}

\section{Introduction}
Gamma-ray bursts are characterized by a high variability.  Their
duration varies from milliseconds to minutes (Norris et al.\
1996)\nocite{1996ApJ...459..393N}. The cosmological distances at which
the bursts occur indicates that the total energy generated (assuming
isotropy) must be at least $10^{52}$\, ergs.  The rapid rise in luminosity and
the variability on short time scales suggest that the radiation is
generated in a region of the size of the order of a hundred kilometers
(a few light milliseconds).  The long duration of some bursts 
indicates that the energy generation within this region
has a rather long time scale.  The brightest gamma-ray bursts are time
asymmetric (Nemiroff et al.\ 1994).\nocite{1994ApJ...435L.133N}  The
complex temporal structure 
of the energy release reflects the activity of a highly variable inner
engine (Fenimore et al. 1996; Sari \& Piran 1997; Kobayashi et al.\
1997).\nocite{sp97}\nocite{kps97}

Short millisecond gamma-ray bursts consist of a single fast rise and
exponential or power-law decay where longer (seconds to minutes)
bursts often show multiple events separated by short time intervals.
These temporal structures are not understood. Explanations range from
multiple shock fronts running into an ambient medium (Sari et al.\
1996),\nocite{1996ApJ...473..204S} expanding shells with brighter
patches and dimmer regions (Fenimore et al.\
1996)\nocite{1996ApJ...473..998F} 
to repeated series of pulses with
Gaussian or power-law profiles (Norris et al.\
1996).\nocite{1996ApJ...459..393N} A clear physical explanation lacks
and the geometry is not at all clear. Combinations of hydrodynamic,
deteriation, expansion and dilation time scales are introduced without
a clear physical explanation or understanding.

Popular models for gamma-ray bursts range from coalescing neutron-star
binaries\, (\cite{1984SvAL...10..177B}), compact objects merging with
the central massive black hole of a galaxy\,
(\cite{1994A&A...290..364R})
collapse of a magnetized white dwarf to a neutron star (Yu \& Blackman
1997)\nocite{1997ApJ...482..383Y} to hyper novae\,
(\cite{1997hgrb.conf..M01P}). All these models have great difficulty
explaining the duration (\cite{1997astro.ph.11354M}) and the intrinsic
variability (Fenimore et al.\ 1996; Sari \& Piran 1997) of the burst.
Roland et al.\
(1994)\nocite{1994A&A...290..364R} proposed that the light curves of
extra galactic gamma-ray bursts can be explained with beamed emission
from a precessing accretion disc around a $10^5$ to $10^6$\,\msun\
black hole. This model was not worked out in greater detail because it
was expected to produce time symmetric light curves with periodicities
on all time scales (Blackman et al.\,1996; Fargion 1998).
\nocite{1998astro.ph..8005F}\nocite{1996ApJ...473L..79B}
Collimated emission, however, received some attention because of its
smaller energy requirements (Hartmann \& Woosley 1995; Dar 1998).  \nocite{1995AdSpR..15e.143H}
\nocite{1998astro.ph..9163D}

Mergers between compact objects have been popular to explain gamma-ray
bursts. The observed burst rate ($\sim 10^{-6}$ events per year per
galaxy, Mao \& Paczy\'nski 1992)\nocite{1992ApJ...388L..45M} is more
than an order of magnitude smaller than estimates for the merger rate
of binary neutron stars (Phinney 1991; Naryan et al.\
1991).\nocite{1991ApJ...380L..17P}\nocite{1991ApJ...379L..17N} This
suggests that the majority of events are hidden. This can be
understood by beaming the emission.  For these sources only the
afterglow would be observable (Rhoads
1997).\nocite{1997ApJ...487L...1R} Beaming of the gamma-ray emission
is also favored because un-beamed models require an enormous amount of
energy generated by distant gamma-ray bursts (Me\'sz\'aros et al.\
1998)\nocite{astro-ph/9808106}.  After glows of gamma-ray bursts
appear to be close to star-forming regions (Paczy\'nski
1998).\nocite{1998ApJ...494L..45P} This un-favors the neutron star
merging model (Bagot et al.\ 1998; Portegies Zwart \& Yungelson 1998).
\nocite{1998A&A...332L..57B}\nocite{1998A&A...332..173P} Mergers
between a black hole and a neutron star do not have this disadvantage
as the space velocities of these binaries is smaller and their merger
time is shorter.

Recently, Portegies Zwart (1998)\nocite{spz98} proposed the gamma-ray
binary as a model for gamma ray bursts.  In such a binary a neutron
star fills its Roche-lobe and transfers mass to a
black hole.  Mass transfer is stable for several
seconds for black hole masses in a small range. If the mass of the
black hole is smaller than $\sim 2.2$\,\msun\ the binary is
dynamically unstable and for black holes more massive than $\sim
5.5$\,\msun\ the system is gravitationally unstable.
For the binaries in which the mass of the black hole falls between
these limits the neutron star material spirals in the black hole 
via an accretion disc.
The magnetic field which is anchored in the disc,
threads the black hole and taps its rotation energy via the
Blandford-Znajek (1977)\nocite{bz77} mechanism. 

This model can explain the complex temporal structure of
gamma-ray bursts in a natural way.  Gamma-rays are emitted in a
collimated cone or beam.  The generated flux within the beam vanishes
at the center of the locus, has largest value near the opening angle
and dimmers to the edge.  Precession of the inner part of the disc
causes the luminosity cone to sweep through space resulting in repeated
pulses or flashes for the observer at a distant planet.  

The next section reviews the proposed model for the 
gamma-ray burst and discusses the expected range of parameters. In \S\
3 the geometry of the model is explained and how a precessing disc
leads to a complex light curve is described.  A classification scheme
is set out in \S\, 4 and we fit the proposed model to a number of
observed gamma-ray bursts. Our findings are discussed in \S\,5.

\section{Intrinsic parameters for the gamma-ray burst}

\subsection{The gamma-ray binary}

The accretion rate onto a neutron star in a common
envelope stage can be highly super Eddington (Chevalier 1993; Brown
1995).\nocite{Che93}\nocite{bro95} The neutron star cannot support
this extra mass and collapses to a black hole.  The result is a close
binary system with a helium star (the remainder of the giant) and a
black hole (the collapsed neutron star). At the end of the
common-envelope phase the mass of the black hole is between
2.4\,\msun\ and 7.0\,\msun\ (\cite{WB96}; \cite{BB98}). The
distribution of black hole masses within this interval is uncertain and depends
strongly on the density distribution in the envelope of the giant and the
duration of the spiral in.

A neutron star is formed after the collapse of the helium core.  The
sudden mass loss and the imparted velocity kick to the nascent
neutron star may dissociate the binary. If the
system remains bound, however, a neutron star -- black hole binary is formed.

The separation between the two stars shrinks due to gravitational wave
radiation (see \cite{PM63}).  When the separation is small enough
(orbital separation $a \aplt 6$\,\rsun) the neutron star fills its
Roche lobe to the black hole within the age of the Universe.  Mass
transfer from the neutron star (with mass $m$) to the black hole (with
mass $M$) is driven by the emission of gravitational waves but
coalescence is prevented by the redistribution of mass in the binary
system.  The entire episode of mass transfer lasts for several seconds
up to minutes. Mass transfer becomes unstable when the mass of the
neutron star drops below the stability limit of $\sim 0.1$\,\msun\,
since the neutron star starts to expand rapidly.  Initially the
material of the neutron star falls in the black hole almost radially
but at a later stage an accretion disc can be formed (see Portegies
Zwart 1998).  This accretion disc can support the strong magnetic
field which threads the black hole and taps its rotation energy via
the the Blandford-Znajek mechanism (Blandford \& Znajek 1977;
MacDonald \& Thorne 1982).\nocite{1982MNRAS.198..345M} The rotational
energy of such a black hole is $\sim 29 \%$ of its mass if maximal
rotating, i.e.; $\sim 10^{54} M/\msun$\,ergs. The luminosity generated 
in the
Blandford-Znajek process is \nocite{bz77}\nocite{tpm86}
\begin{equation}
	L \approx 10^{51} \left( {\mu M \over 3 [\msun]} \right)^2
	                  \left( {B \over 10^{15} [G]} \right)^2 
	  \;\; [{\rm erg\,s}^{-1}].
\end{equation}
Here $\mu$ is the angular momentum of the black hole relative to that
in maximal rotation.  A strong magnetic field ($B \sim 10^{15} G$) is
required to drive the engine. How it is generated is not well
understood. However, strong magnetic fields in black holes have gained
a lot of support over the last few years (see e.g.\ Paczy\'nski 1998b).
Also in proto neutron stars (Pal et
al. 1998)\nocite{astro-ph.9806356P} and soft gamma-ray repeaters
(Kouveliotou et al.\ 1998)\nocite{1998Natur.393..235K} strong magnetic
fields of the order of $10^{15}$\,Gauss are favored.  The magnetic
field is also anchored in the disc but the luminosity comes from the
spin energy of the black hole (Katz 1997).\nocite{1997ApJ...490..633K}

\subsection{Precession of the accretion disc}

The misalignment in the spin axis of the black hole and the angular
momentum axis of the binary causes the accretion disc around the
black hole to precess.  Rigid body precession of such a Keplerian disc
is possible if the sound crossing time scale 
of the disc is small compared to the precession time scale (Papaloizou
\& Terquem 1995; Larwood et al.\ 1996), which is generally the case.
\nocite{1995MNRAS.274..987P}\nocite{1996MNRAS.282..597L}
The forced precession period $\tau_{\rm pre}$ is then
(Larwood 1998)\nocite{Larw98}
\begin{equation}
\tau_{\rm pre} = \frac{7}{3} P_{\rm orb} {(1+q)^{1/2} \over q \cos \nu} 
	    \left( \frac{a}{r_{\rm disc}} \right)^{3/2} \;\;\;  \unit{s}.
\label{Eq:Ppre}\end{equation}
Here $\nu$ is the angle between the orbital angular momentum axis and
the spin axis of the accreting star.  The mass ratio $q \equiv m/M$,
the orbital period $P_{\rm orb}$ and the filling factor of the
accretion disc around the black hole $a/r_{\rm disc}$, where $a$ is
the semi-major axis of the orbit and $r_{\rm disc}$ is the size of the
accretion disc, change in time as the neutron star loses mass to the
black hole. 

At any time an upper limit to the size of the accretion disc is given
by the size of the Roche-lobe of the black hole $R_{\rm Rl}$.  For
example: a 1.4\,\msun\ neutron star fills its Roche lobe to a
3\,\msun\ black hole at an orbital separation of about 42.7\,km. After
1.15\,\msun\ of the neutron star material is transferred conservatively
to the black hole, an accretion disc starts to form (see Portegies
Zwart 1998). At this moment the
orbital separation has widened to approximately 135\,km (assuming that
the neutron star is prescribed by a Newtonian polytope), and $a/R_{\rm
Rl} \approx 1.21$.  For small mass ratio the size of the accretion
disc is approximately 0.87 times the Roche-radius of the accretor
(Paczy\'nski 1977),\nocite{pac77} i.e.: $a/r_{\rm disc} \approx 1.39$.

Substitution of the parameters from the example into Eq.\ref{Eq:Ppre} 
results in a lower limit to the precession period of the disc at the
moment of its formation
\begin{equation}
\tau_{\rm pre} \geq \frac{0.94}{\cos \nu} \;\;\;   \unit{s}.
\label{Eq:Pp}\end{equation}

As mass transfer proceeds the precession period may change. In the
model we neglect this effect but we return to it in the discussion
\S\,\ref{Sect:discussion}. 

\subsection{Misalignment of the angular momentum axis}

Upon the formation of the neutron star the asymmetry in the supernova and
the accompanying velocity `kick' may cause the orbital angular
momentum axis to change direction, because the kick will generally be
somewhat out of the orbital plane of the binary.

The angular momentum axis of the binary after the supernova makes an
angle $\nu$ with that before the supernova. If we
assume that mass lost in the supernova event does not affect the
orbital angular momentum axis and the nascent neutron star receives a
velocity kick of magnitude $v_{\rm k}$ in the direction $\thetak$ (the
angle with the orbital plane) and $\phik$ (the angle between the
direction of motion of the helium star and the kick velocity in the
orbital plane) the new angular momentum axis follows from simple
geometry:
\begin{equation}
	\cos \nu  = \frac{1 + \tilde v \cos \thetak \cos\phik}
		         {[\tilde v^2 \sin^2\thetak 
                          + (1 + \tilde v \cos \thetak \cos \phik)^2]^{1/2}}
\label{Eq:cos_nu}\end{equation}
Here $\tilde v \equiv v_{\rm k}/v_{\rm orb}$, with $v_{\rm orb}$ the
relative orbital velocity of the two stars prior to the supernova.

We can calculate the probability distribution for the misalignment
angle $\nu$ from the binary parameters before the supernova and by
selecting a velocity distribution for the kick.
The magnitude of the velocity kick is taken randomly from
the distribution proposed by Hartman
(1997)\nocite{1997A&A...322..127H} as the intrinsic velocity
distribution for single radio pulsars. 
This distribution is flat at
velocities below 250\,\kms\ but has a tail extending to several
thousand of \kmsp  
\begin{equation}
P(u)du = {4\over \pi} \cdot {du\over(1+u^2)^2},
\label{Eq:kick}\end{equation}
with $u=v/\sigma$ and $\sigma = 600$~\kmsp
The resulting probability distribution for the misalignment angle
$\nu$ is presented in Fig.\,\ref{Fig:Pnu} for various binary parameters.

Figure\,\ref{Fig:Pnu} shows that the majority of the
binaries have $\nu \aplt 20^{\circ}$, although larger angles are not
excluded. A small fraction ($\aplt 1\%$) of the binaries will even
have retrograde angular momentum axes. 

If the spin axis of the black hole is aligned with the angular
momentum axis of the binary before the supernova, Fig.\,\ref{Fig:Pnu}
gives the probability distribution for the angle between the post
supernova angular momentum axis and the spin axis of the black hole.
Substitution of the average angle between the angular momentum axis of
the binary and the spin axis of the black hole into Eq.\,\ref{Eq:Pp}
gives a precession period of about a second.

\placefigure{Fig:Pnu}

\section{The precessing gamma-ray jet}

A lantern mounted along the rotation axis of a precessing object
projects circles of light in a regular pattern. Nutation of
the object causes the light cone to make a number of small circles
within each larger precession circle. 

The accretion disc in the model precesses around the black
hole. 
The magnetic field is anchored in the disc and precesses with
the same period.
The radiation cone therefore also precesses with the same period.

The rotation period of the black hole is much smaller than the
precession period of the disc.  But even if the radiation cone would
have the same period as the black hole this could not be observed
with todays detectors due to their lack of time resolution.
We return to this in the discussion \S\,\ref{Sect:discussion}

\subsection{The precessing locus}

Slaved disc precession is a well known phenomenon in X-ray binaries
(van den Heuvel et al.\ 1980; Hut \& van den Heuvel 1981; Band \&
Grindlay
1984)\nocite{1980A&A....81L...7V}\nocite{1981A&A....94..327H}\nocite{bg84}
and is also used to describe the complex light curves of extra galactic
radio sources (Falcke \& Biermann 1998).\nocite{1998astro.ph.10226F}

We use the same principle of the slaved disc around the
accreting black hole to explain the complex light curves of bright
gamma-ray bursts. The polar angle $\theta$ and
the azimuth angle $\phi$ are defined in a spherical polar coordinate
system.  The spin axis of the black hole (the locus of the gamma-ray
jet) can then be parameterized with $\thetaj$ and $\phij$ (Band \&
Grindlay 1984)
\begin{eqnarray}
\phij &=& \Omega_{\rm pre} (t-t_{\rm \circ}) +
    \frac{\Omega_{\rm pre}}{\Omega_{\rm nu}}\sin ( \Omega_{\rm nu} t
    ), \nn
\thetaj &=& \theta^{\rm \circ}_{\rm jet} +
    \frac{\Omega_{\rm pre}}{\Omega_{\rm nu}}\tan\theta^{\rm \circ}_{\rm jet} 
		\cos ( \Omega_{\rm nu} t ).
\end{eqnarray}
Here the precession and nutation frequencies are
$\Omega_{\rm pre} = 2\pi/\tau_{\rm pre}$, 
$\Omega_{\rm nu}  = \pm 2\pi/\tau_{\rm nu}$
with both periods $\tau_{\rm pre}$ and $\tau_{\rm nu}$ of the order of
a second (see Eq.\,\ref{Eq:Pp}). 
The direction of the observer can also be parameterized
using $\thetao$ and $\phio$.
The angle $\psi$ between the observer $\hat r_{\rm obs} (\thetao, \phio)$
and the central locus of the jet $\hat r_{\rm jet} (\thetaj, \phij)$ is 
given by $\psi = \cos^{-1}(\hat r_{\rm obs}\cdot\hat r_{\rm jet})$.

\subsection{The intensity distribution within the jet}

A possible mechanism to generate a collimated gamma-ray jet from an
accreting black hole is provided by the Blandford-Znajek process.  In
this process the rotation energy can be extracted from
black hole by slowing down its rotation via the magnetic field which
exert torque on the induced current on the black hole horizon (Thorne
et al.\ 1986).\nocite{tpm86} The torque exerted by a surface element
of the horizon is proportional to the product of magnetic flux $d\Psi$
through the surface element and the current $\I$.  The power of this
torque is carried by the Poynting flux $\Pi$ along the magnetic field
lines, which rotate with angular velocity $\Omega_F$ around the
rotating black hole ($\Omega_H > \Omega_F$).  The Poynting flux
carried out through the magnetic surface between $\Psi$ and $\Psi +
d\Psi$ is (Thorne et al.\ 1986)
\begin{eqnarray} 
d\Pi = \frac{\Omega_F}{2\pi} \I(\Psi) d\Psi.\label{dpi} 
\end{eqnarray} 
Here the magnetic surface is characterized by the embedded magnetic
flux $\Psi$.  The luminosity produced by the Blandford-Znajek process
(Eq. 1) is obtained by integrating Eq.\,\ref{dpi} with average
magnetic field $B$ on the horizon.

Collimation of the Poynting flux is computed for two dimensional
force-free solutions of the stream equation (Fendt 1997). Identifying
the Poynting flux as luminosity $L$ we parameterize the luminosity
distribution in cylindrical coordinates
\begin{eqnarray} 
L(\psi) \equiv L(x) & = & \Omega_F(x) \I(x) \frac{d\Psi(x)}{d x}.
\end{eqnarray} 
Here $x = R/R_L$\ is the cylindrical radius (the distance from the
black hole) normalized to the asymptotic light radius
$R_L=c/\Omega_F$. The angle dependence is obtained from $x = r_f
\sin\psi$ in which $r_f$ is the freeze out distance (from the black
hole and normalized by $R_L$) where the photons can leave freely. The
opening angle $\psi_{\rm open}$ is fixed by $\sin\psi_{\rm
open}\approx 1/r_f$.  A small opening angle $\psi_{\rm open} = 6^{\rm
\circ}$ with $r_f=10$ are assumed.\footnote{The approximation is in
the limit of an observer at infinity.}  The magnetic flux $\Psi$,
current distribution $\I$ and the angular velocity of the magnetic
field $\Omega_F$ are then (Fendt 1997)\nocite{fen97}

\begin{eqnarray}
\Psi(x) &=& \frac{1}{\beta}\ln
\left(1+\left(\frac{x}{\alpha}\right)^2\right), \nn 
\I (\Psi) &=& B \left( 1-e^{-\beta \Psi} \right) e^{-\delta \Psi}, \nn
\Omega_F^2 (\Psi) &=& \frac 14 \gamma \beta^2 B^2 e^{-2\delta\Psi}
-\frac{1}{\alpha^2}, 
\label{Eq:L_BZ}\end{eqnarray}
where 
\begin{eqnarray}
B &\equiv& \left( 1-e^{-\beta} \right)^{-1} e^\delta, \nn
x_{\rm jet} &=& \alpha \sqrt{e^\beta-1}, \nn
\alpha^2 &=& \left( \left(\frac 14 \gamma \beta^2
B^2\right)^{1/(1+2\delta/\beta)}-1\right)^{-1}, \nn 
 \gamma &=& \frac{4}{\beta^2} \left( 1-e^{-\beta} \right) ^2e^{-2\delta}
       \left(1+\frac{e^\beta-1}{x_{\rm jet}^2}\right)^{1+2\delta/\beta}.
\end{eqnarray}
The following parameters are used: $\alpha=0.5$, $\beta=6$ and
$\delta=0.3$.  Fig.\,\ref{Fig:Lphi} presents the $\psi$ dependence of
the luminosity for a range of parameters.  Note that the luminosity
vanishes at $\psi=0$ and has a peak near the opening angle, which are
characteristics of this distribution. 

%
%--------------------------------------------------
%Assuming the uniform $\Omega(x)$, we get the luminosity per unit area
%\be
%L(x) &\propto& \frac{1}{1+(\frac{x}{d})^2}
% \frac{1-e^{-b \Psi(x)}}{1-e^{-b}}.
%\ee
% For the simplest, we assume that the luminosity
% decreases exponentially for $\psi>\psi_{open}$ ($x>1$)
%\be
%L(x>1) &\propto& \exp(-(\psi-\psi_{open})/\psi_0)
%\ee
%with $\psi_0=0.2\times \psi_{open}$.
%--------------------------------------------------

\placefigure{Fig:Lphi}

\subsection{Time dependency of the intrinsic luminosity}

The intrinsic time variation of a single gamma-ray burst has a short
rise time followed by a linear decay (Fenimore
1997).\nocite{astro-ph/9712331} We construct a burst time
profile from three components: an exponential rise with
characteristic time scale $\tau_{\rm rise}$, a plateau phase with
time scale $\tau_{\rm plat}$ and a stiff decay with
time scale $\tau_{\rm decay}$.
We selected the following function:
\begin{equation}
                %I (t) = N_I \left( \frac{t}{\tau}
                %\left(1-\frac{t}{t}\right)^2\right)^{0.25} 
I(t) = N_I \left( 1-e^{-t/\tau_{\rm rise}} \right) \left( \frac{\pi}{2}-
   \tan^{-1} \left[ (t-\tau_{\rm plat})/\tau_{\rm decay} \right]
  \right).
\end{equation}
Here $I(t)$ is normalized with constant $N_I$ to fit $I(t)=1$ at its 
maximum.
Fig.~\ref{Fig:Iintrinsic} gives an example of the function for the
selected parameters.

\placefigure{Fig:Iintrinsic}

The choice of this function is rather arbitrary. Lack of understanding
of the inner engine which drives the burst provides the freedom to
choose this profile. The main reason to 
select this function is that it is smooth and easy to adjust with three
time scales; a rise time, a plateau time and a decay time.

\section{The light curves}

We illustrate the light curves produced by the model
in \S\,\ref{Sect:examples}. Fits to real observed
gamma-ray bursts are presented in \S\,\ref{Sect:fits}.

\subsection{Examples of light curves}\label{Sect:examples}

To study the general behavior of the temporal structure of the
model we define the frequency ratio, 
\begin{eqnarray}
	R_\Omega\equiv\Omega_{\rm nu}/\Omega_{\rm pre},
\end{eqnarray}
which can be negative.

To illustrate the temporal behavior of the intrinsic intensity we select
$\tau_{\rm rise}=0.2 \tau$, $\tau_{\rm plat}=5 \tau$ and $\tau_{\rm
decay}=\tau$. 

In the following series of figures we show how the behavior of
the temporal structure of the luminosity curve changes by varying the
observation angle $\thetao$ and $\phio$, the initial direction of the
spin axis of the locus $\theta^{\rm \circ}_{\rm jet}$, 
the nutation frequency relative to the precession
frequency $R_\Omega$ and the precession period $\tau_{\rm pre}$.  For
simplicity the selected parameters are 
presented in the following format ($\tau_{\rm pre}$,
$R_\Omega$, $\theta^{\rm \circ}_{\rm jet}$; $\thetao$ $\phio$).
Since we do not yet attempt to fit any data we chose $t_{\rm pre} =
\tau$, $t_{\rm \circ} = 0$ and $\phio=0$ for most examples.

\placefigure{Fig:0_1_02_0_10_0}

Figure\,\ref{Fig:0_1_02_0_10_0} gives the observed light curve for a
fixed angle $\thetao = 10^{\rm \circ}$. The luminosity cone is
steady rotating, the observer makes an angle of $10^{\rm \circ}$
with the central locus of the luminosity cone, in this case the spin axis of
the black hole. The result is an observed flux smaller than at its
maximum but the luminosity curve has the same temporal shape as
Fig.\,\ref{Fig:Iintrinsic}: an angle of $\thetao \approx 4^{\circ}$
would result in Fig.\,\ref{Fig:Iintrinsic}.  
The observers line of sight moves clockwise along the solid curve in
the upper right corner of Fig.\,\ref{Fig:0_1_02_0_10_0}.  Increasing
$t_{\rm \circ}$ results in a counter clockwise shift of the starting
point indicated with the $\bullet$ in Fig.\,\ref{Fig:0_1_02_0_10_0}.

\placefigure{Fig:0_1_02_2_0_0}

Figure\,\ref{Fig:0_1_02_2_0_0} demonstrates the effect of varying only 
three of the parameters while  $\thetao = 0$. Increasing $\theta^{\rm
\circ}_{\rm jet}$ causes the precession and nutation of the luminosity 
cone to make larger circles in space. Sometimes our line of sight
completely misses the cone which results in the absence of flux. Increasing
$\theta^{\rm \circ}_{\rm jet}$ to about $90^{\rm \circ}$ results in a
picket fence pattern as the luminosity is pulsed with high frequency
(not shown).

\placefigure{Fig:0_1_02_2_12_0}

Figure\,\ref{Fig:0_1_02_2_12_0}  %% and \ref{Fig:0_1_02_2_2_0} 
shows the effect of changing $R_\Omega$ and $\theta^{\rm \circ}_{\rm jet}$.
The pattern becomes irregular and by changing the parameters the
pattern quickly changes. 

%\begin{figure}[ht]
%\centerline{\psfig{file=Ft0_1_02_2_2_0.ps,height=4cm,angle=-90}}
%\centerline{\psfig{file=Ft0_1_01_2_2_0.ps,height=4cm,angle=-90}}
%\centerline{\psfig{file=Ft0_2_07_12_10_0.ps,height=4cm,angle=-90}}
%\centerline{\psfig{file=Ft0_2__07_12_10_0.ps,height=4cm,angle=-90}}
%\centerline{\psfig{file=Ft0_2_06_12_12_185.ps,height=4cm,angle=-90}}
%\caption{From upper to 
%	lower panel gives the burst profile with parameters 
%	($t_{\rm \circ}$=0; 
%	$\tau_{\rm pre}$=1, $R_\Omega$=0.2;
%	$\theta^{\rm \circ}_{\rm jet}$=2, $\thetao$=2; 
%	$\phio$=0) for	the upper panel followed by 
%	(0; 2, 0.1; 2, 2, 0), 
%	(0; 2, 0.7; 12, 10, 0),
%	(0; 2, -0.7; 12, 10, 0) and
%	(0; 2, 0.6; 12, 12, 185) respectively.   
%}
%\label{Fig:0_1_02_2_2_0}
%\end{figure}

\subsection{Fitting}\label{Sect:fits}.

The model contains a total of nine parameters
(Tab.\,\ref{Tab:param}) which are more or less free to choose within
the theoretical framework. For practical reasons we introduced the dead
time $\tau_{\rm dead}$ which is the initial time between the BATSE
trigger and the start-up of the real burst. In most cases this
time scale is of the order of a second or so.
The intrinsic parameters for the $\Psi$-dependence of the intrinsic
luminosity within the cone ($\alpha$, $\beta$ and $\delta$)
are treated as fixed parameters.

% Table: Stellar evolution
\placetable{Tab:param}

Fitting a nine-parameter model is not a trivial exercise, especially
if the result depends sensitively on their proper values. 
Even if fitting is possible at all it is still not certain what the fitted
model tells us about the underlying physical process. In some cases
equally satisfactory fits are obtained with different parameters.

The first step in the fitting procedure is to determine the
background. This is done on the initial, $\sim 1800$, time bins of
64\,ms in the data stream of each observed burst. The average
count rate in this part is used as background.
Note that modeling the return to the background at the end of
the burst is as important as modeling the peaks of the bursts. 

We use simulated annealing (Press et al. 1992)\nocite{NumResp} to
fit the observed bursts. The parameters described in
Tab.\,\ref{Tab:param} are chosen freely but initially preferred values
are suggested. After each iteration we determine the $\chi^2$ from
the fit of the binned (in 64\,ms bins) model data to the observed
burst profile.  This value is minimized by the annealing
algorithm. The temperature in the process was reduced according to $T
= T_{\circ}(1-k/K)^4$. Here $T_{\circ}$ is the highest value of
$\chi^2$ found in the initial set of nine randomly selected bursts,
$k$ is the cumulative number of iterations and $K$ is the scheduled
number of iterations (in most cases between $10~4$ and $10^5$. The
temperature was lowered after each 900\, iterations.

We applied this method to the same sample of gamma-ray bursts which
was used by Norris et al.\ (1996) to fit flattened Gaussians to
individual pulses in complex bursts profiles. We fitted the energy
channel between 115\,keV and 320\,keV, which has the highest counts
for most bursts and is therefore least plagued by noise (the fitting
procedure has the tendency to fit the noise as well).  The following
set of figures give the observed bursts (in the upper panels) and the
result of the fit to each burst in the lower panel. The geometry of
the system is given in the upper right corner of the lower panel. The
parameters used for the fit are also presented in
Tab.\,\ref{Tab:fits}. The values of $\chi^2$ where we decided to stop
the fitting routine are not given.

% Table: fitted GRBs
\placetable{Tab:fits}

Table \,\ref{Tab:fits} and the Figs.\,\ref{Fig:GRB143} to
\ref{Fig:GRB2228} give the results of fits to a number of gamma-ray
bursts. The peak intensity in the modeled bursts (see $I_{\rm max}$ in
Tab.\,\ref{Tab:fits}) is generally close to the intrinsic maximum of
the burst but the observable fluency ($F_{\rm int}$) is generally much
smaller than the intrinsic integrated luminosity of the burst.

Figures\,\ref{Fig:GRB143} and \ref{Fig:GRB543} give the fits to BATSE
trigger number 143 and 543, respectively. Only the overall shape of
the bursts are fitted reasonably well. Smaller details appeared to be
hard to fit with the selected parameters.

\placefigure{Fig:GRB143}
\placefigure{Fig:GRB543}

Figures\,\ref{Fig:GRB999}, \ref{Fig:GRB1425} and \ref{Fig:GRB1609}
gives the result of the fits to BATSE trigger numbers 999, 1425 and
1609, respectively.  The result of these fits are rather
satisfactory. Even the individual peaks within each burst
tend to be asymmetric which is also the case in real gamma-ray bursts.

\placefigure{Fig:GRB999}
\placefigure{Fig:GRB1425}
\placefigure{Fig:GRB1609}

Figures\,\ref{Fig:GRB1683} and \,\ref{Fig:GRB1683A}, BATSE trigger
number 1683, shows two fits to a very complicated burst (see also
Tab.\,\ref{Tab:fits}). The fitting procedure produces several
solutions which a broad range of shapes ranging from a single smooth
curve to sharp symmetric peaks.

\placefigure{Fig:GRB1683}
\placefigure{Fig:GRB1683A}

Possibly not all peaks in BATSE trigger number 1974,
Fig.\,\ref{Fig:GRB1974}, are real. 
We decided to present the smooth solution which 
reproduces the over all shape of the burst.

\placefigure{Fig:GRB1974}

BATSE trigger 2067, Fig.\,\ref{Fig:GRB2067}, shows a lot of small
details of which some may be noise. In this case we decided to
show the fit with a lot of small structure to demonstrate how the model 
attempts to reproduce the noise if the adopted fitting procedure is
continued for too long.

\placefigure{Fig:GRB2067}

The observed burst with BATSE trigger 2228, Fig.\,\ref{Fig:GRB2228}, is
extremely spiky. We show the smooth fit to the burst which presents the
overall shape but neglects the smaller details. In this burst we had
difficulty reproducing the smaller structures as the fitting procedure 
attempts to fit the peaks more accurately than the valleys.

\placefigure{Fig:GRB2228}

\section{Discussion}\label{Sect:discussion}

The relatively simple geometry of a precessing accretion disc around a
black hole provides an excellent model for long lasting gamma-ray
bursts with a complex temporal structure. The large number of free
parameters and the low quality of the observational data, however,
make it hard to fit the observations and to draw conclusions on the
resulting parameters. There are often several solutions which lead to
equally satisfactory fits but at a different total luminosity (compare
Figs.\,\ref{Fig:GRB1683} and \ref{Fig:GRB1683A}). This makes
classification difficult.

Observed gamma-ray bursts are asymmetric at all time scale (Norris et
al.\ 1996). Some fitted examples, however, show symmetric peaks, which
limits the reliability of the model and the fits (Nemiroff
et al.\ 1994). 
On the other hand, we assumed that the precession and nutation period,
and their amplitudes are all constant as in a force free environment. Also
the black hole is assumed to rotate around its center of mass, which
is not affected by variations in the binary parameters, or the
accretion disc.  Including these parameters complicates the model
considerably and may cause individual peeks to become asymmetric.

The advantage of our model is that it follows rather natural 
from a system in which a black hole is orbited by a precessing
accretion disc.
Precessing accretion discs are common phenomena and the temporal
structure of X-ray binaries are often explained with a similar model
(see Larwood 1998 and references herein).\nocite{Larw98}
Apart from the similarity with existing objects like X-ray binaries
and active galactic nuclei the precessing beam solves two problems: 1)
the irregular behavior of the precessing beam can cause the burst to
show up and disappear at unpredictable moments. 2) the radiation is
not isotropic and therefore the amount of energy radiated is
considerably smaller than the estimate based on the assumption of
spherical symmetry (see third column in Tab.\,\ref{Tab:fits}).  The
energy budget of observed gamma-ray bursts reflects then a
considerable fraction of the total amount of energy radiated
(M\'esz\'aros et al. 1998).\nocite{astro-ph/9808106} 

The characteristic energy dependence of observed gamma-ray bursts (the
peaks get narrower for higher energies) could be explained by a
decrease in the opening angle with energy. However, lack of the
background physics prevents us from modeling this process in greater
detail.
 
Another interesting implications is that all gamma-ray bursts may
intrinsically produce the same amount of energy but since only a small 
fraction of the produced energy is detectable they cannot be used as
standard candles.

If the inner engine would be able to work for a longer period, 
days or weeks, repeated bursts from the same source may be observed
within these time intervals. Mass transfer from a neutron star to a
black hole is not likely to last for more than a few minutes but if a
similar binary with a white dwarf as donor could produce similar
characteristics the time scale on which the inner engine is working 
becomes of the order of hours or days.

We assumed that the spin axis of the black hole and the locus of
the gamma-ray jet are aligned. The introduction of an extra angle
would strongly alter the $\psi$-dependence of the luminosity generated
in the jet. In that case the gamma-ray jet passes our line of sight
with the spin frequency of the black hole. This time scale is much
shorter than the time resolution of the current gamma-ray
detectors. A significant improvement in the time resolution of
gamma-ray detectors may find such objects to be visible as gamma-ray
pulsars. The pulse period in such pulsars will be of the order of the
rotation period of the black hole.

Within the discussed model for a gamma-ray burst from a gamma-ray
binary three configurations can be imagined which all lead to a
different observational phenomenon:
1) Gamma-ray bursts with one or more peaks in the line of the
model discussed in this paper.
2) Systems where mass transfer from the neutron star to the black
hole is unstable and directly results in a merger (see Portegies Zwart 
1998).
3) A short burst if the black hole spins retrograde relative to
the binary angular momentum axis. This happens in less than 1\% of the bursts.
Observations also distinguish between three different classes of
gamma-ray bursts: short and faint bursts, long
and bright and a third class of intermediate bursts\,(Mukherjee et
al.\ 1998).\nocite{astro-ph/9802085} 

%% extra citation within the figure captions.
%% Due to funny ApJ style files unable to put the directly in the 
%% figure caption.....
\nocite{1997MNRAS.291..569H}

\acknowledgments SPZ is grateful to Gerry Brown for inviting him for
an extended visit to the University of Stony Brook, and Jun Makino for 
discussions and critically reading the manuscript. We thank the
anonymous referee for his critical and helpful comments.
This work was supported in part by the U.S. Department of Energy under
Grand No. DE-FG02-88ER40388. HKL is supported also in part by KOSEF-985-
0200-001-2.

%\bibliographystyle{/home3/spz/tex/lib/inputs/aabib}
%\bibliography{/home3/spz/tex/lib/inputs/references,/home3/spz/tex/lib/inputs/spz}

\clearpage

\figcaption[M3m4a0536ppMxw.ps]
{ Cumulative probability distribution of angles $\nu$ in
	degrees (X-axis).  The progenitor system is assumed to contain
	a 3\,\msun\ black hole and a 4\,\msun\ helium star which
        transforms into a 1.4\,\msun\ 
	neutron star in the supernova. A kick with a magnitude given
	by Eq.\,\ref{Eq:kick} (solid lines) in a random direction is
	imparted to the neutron star. The separation of the circular
        orbit before
	the supernova was taken to be 0.5\,\rsun, (uppermost left
	curve), 3\,\rsun\ (for the middle line) and 6\,\rsun\ (lower
	right line). The dotted lines give the cumulative probability
	distribution if the kick was selected randomly from a
	Maxwellian with a three-dimensional velocity dispersion of
	270\,km/s (see Hansen \& Phinney 1997). 
\label{Fig:Pnu}
}

\figcaption[lum_psi4.ps]
{	$\psi$ dependence of the Luminosity from a few sets of
        parameters according to Eq.\,\ref{Eq:L_BZ}. Each curve is 
        normalized to unity at its maximum: 
        $\alpha=0.5$, $\beta=6$ and $\delta=0.3$ for the solid line,
        which we used throughout this paper,
	$\alpha=0.6$, $\beta=6$ and $\delta=0.3$ (dashed line),
        $\alpha=0.5$, $\beta=3$ and $\delta=0.3$ (dotted line)
        and $\alpha=0.5$, $\beta=6$ and $\delta=0.4$ for the
        dash-dotted line.
\label{Fig:Lphi}
}

\figcaption[lum_int.ps]
{Time dependence of the intrinsic luminosity
          with $\tau_{\rm rise}=0.2 \tau$, $\tau_{\rm plat}=5 \tau$,
          and $\tau_{\rm decay}=1 \tau$, where $\tau$ is in arbitrary
          time units.
\label{Fig:Iintrinsic}
}

\figcaption[Ft0_1_02_0_10_0.ps]
{The flux (Y-axis) as a function of time (X-axis) for the
	parameters ($\tau_{\rm pre}$=1,
	$R_\Omega$=0.2, $\theta^{\rm \circ}_{\rm jet}$=0;
	$\thetao$=10, $\phio$=0).  The upper right corner gives a
	schematic representation of the central locus of the black
	hole (central $\circ$) and the trajectory of our line of sight
	(thick solid line) starting at the $\bullet$, moving
	clockwise. The inner dotted line identifies the angle at which
	the luminosity distribution of Fig.\,\ref{Fig:Lphi} is
	maximum, it drops to zero at the outer dotted line. 
\label{Fig:0_1_02_0_10_0}
}

%\figcaption[Ft0_1_01_2_0_0.ps]{}
%\figcaption[Ft0_1_02_4_0_0.ps]{}
%\figcaption[Ft0_4_02_2_0_0.ps]{}
%\figcaption[Ft0_4_4_12_0_0.ps]{}
\figcaption[Ft0_1_02_2_0_0.ps]
{The observer looks directly in the spin axis of the black hole
	and $\theta^{\rm \circ}_{\rm jet}$, $R_\Omega$ and $\tau_{\rm
	pre}$ are varied.  The selected parameters are
	($\tau_{\rm pre}$=1,
	$R_\Omega$=0.2, $\theta^{\rm \circ}_{\rm jet}$=2;
	$\thetao$=0, $\phio$=0) for the upper panel followed by
	(1, 0.1, 2; 0, 0), 
	(1, 0.2, 4; 0, 0), 
	(4, 0.2, 2; 0, 0) and 
	(4, 4, 12;  0, 0) for the lower panels, respectively.  
\label{Fig:0_1_02_2_0_0}
}

%\figcaption[Ft0_1_01_2_12_0.ps]{}
%\figcaption[Ft0_1_02_4_12_0.ps]{}
%\figcaption[Ft0_4_02_2_12_0.ps]{}
%\figcaption[Ft0_4_4_12_12_0.ps]
\figcaption[Ft0_1_02_2_12_0.ps]
{From upper to
	lower panel gives the burst profile with parameters 
	($\tau_{\rm pre}$=1, $R_\Omega$=0.2,
	$\theta^{\rm \circ}_{\rm jet}$=2; $\thetao$=12, 
	$\phio$=0) for	the upper panel followed by 
	(1, 0.1, 2; 12, 0), 
	(1, 0.2, 4; 12, 0), 
	(4, 0.2, 2; 12, 0) and
	(4, 4,  12; 12, 0) respectively.   
\label{Fig:0_1_02_2_12_0}
}

\figcaption[GRB143E3.ps] 
{ 	The Gamma-ray burst with BATSE trigger 143. 
	The upper panel gives the observed burst, the lower panel the fitted
	model with an impression of the geometry in the right corner of the
	lower panel.
Time along the X-axis is in units of 64\,ms, the time resolution of
the BATSE detector. The number of counts is along the vertical axis.
The fit parameters are given in
Tab.\,\ref{Tab:fits}.  
\label{Fig:GRB143}
}

\figcaption[GRB543E3.ps]
{BATSE trigger 543. 
\label{Fig:GRB543}
}

\figcaption[GRB999E3.ps]
{BATSE trigger 999. 
\label{Fig:GRB999}
}

\figcaption[GRB1425E3.ps]
{BATSE trigger 1425. 
\label{Fig:GRB1425}
}

\figcaption[GRB1609E3.ps]
{BATSE trigger 1609. 
\label{Fig:GRB1609}
}

\figcaption[GRB1683E3II.ps]
{BATSE trigger 1683. 
\label{Fig:GRB1683}
}

\figcaption[GRB1683E3A.ps]
{Smooth version of BATSE trigger 1683. 
\label{Fig:GRB1683A}
}

\figcaption[GRB1974E3.ps]
{BATSE trigger 1974.
\label{Fig:GRB1974}
}

\figcaption[GRB2067E3.ps]
{BATSE trigger 2067. 
\label{Fig:GRB2067}
}

\figcaption[GRB2228E3.ps]
{BATSE trigger 2228.
\label{Fig:GRB2228}
}

\clearpage

\begin{deluxetable}{l|l}
\tablecaption{ 
Free parameters which may vary per burst.
}
\tablewidth{0pt}
\tablehead{ \colhead{parameter} & \colhead{note} }
\startdata
$\tau_{\rm dead}$ & initial epoch without signal \\
$\tau_{\rm rise}$ & burst start-up time \\
$\tau_{\rm plat}$ & burst plateau time \\
$\tau_{\rm decay}$& burst decay time \\
$\tau_{\rm pre}$  & precession period \\
$\tau_{\rm nu}$        & nutation period \\
$\theta^{\circ}_{\rm jet}$& precession angle \\
$\theta_{\rm obs}$& observer angle \\
$\phi_{\rm obs}$  & observer angle \\ \hline
\enddata
\label{Tab:param}
\end{deluxetable}

\begin{deluxetable}{lrlr|lrrlllrrr} 
%\begin{table*}
\tablecaption{ 
Fitted parameters for a number of the bursts. The first column gives
the BATSE trigger number for the burst followed by the maximum
observable luminosity as fraction of the maximum luminosity of the
smulated burst, the integrated flux
$F_{\rm int}$ as fraction of the totally emitted radiation
(in percentage). The third column gives the figure number which depicts 
the burst profile. The following nine columns give the fit
parameters (see Tab.\,\ref{Tab:param}).
}
\tablewidth{0pt}
%\colhead{GRB}&
%\colhead{$I_{\rm max}$} &
%\colhead{$F_{\rm int}$}&
%\colhead{Fig.}&
%\colhead{$\tau_{\rm dead}$} & 
%\colhead{$\tau_{\rm rise}$} & 
%\colhead{$\tau_{\rm plat}$} & 
%\colhead{$\tau_{\rm decay}$}& 
%\colhead{$\tau_{\rm pre}$}  & 
%\colhead{$\tau_{\rm nu}$}        & 
%\colhead{$\theta^{\circ}_{\rm jet}$}& 
%\colhead{$\theta_{\rm obs}$}& 
%\colhead{$\phi_{\rm obs}$}  \\
%}
%\begin{flushleft}
%\begin{tabular}{lrlr|lrrlllrrr} \hline\hline
\tablehead{ 
GRB&$I_{\rm max}$&$F_{\rm int}$&Fig.&$\tau_{\rm dead}$ & 
$\tau_{\rm rise}$ & 
$\tau_{\rm plat}$ & 
$\tau_{\rm decay}$& 
$\tau_{\rm pre}$  & 
$\tau_{\rm nu}$        & 
$\theta^{\circ}_{\rm jet}$& 
$\theta_{\rm obs}$& 
$\phi_{\rm obs}$  \\ 
&\multicolumn{2}{c}{[\%]}&[\#]&\multicolumn{6}{c}{[seconds]} &
\multicolumn{3}{c}{[degrees]} } %%\hline
\startdata
143 &98&0.33&\ref{Fig:GRB143}& 2.52&0.31 & 4.00&0.81& 3.22& 0.94&  6.5&  2.9&128.3 \\
543 &98&0.24&\ref{Fig:GRB543}& 1.86&0.15   & 3.94&0.50&10.8 &-26.2&  5.0& 10.1&228.5 \\  
999 &85&0.05&\ref{Fig:GRB999}& 0.92& 0.22  & 5.81&0.50& 5.10&-29.1&  8.6& 17.9&115.2 \\
1425&97&0.28&\ref{Fig:GRB1425}&1.00&18.8   & 8.00&0.20& 1.40& 0.26&  5.1&352.8&139.0 \\
1609&80&0.17&\ref{Fig:GRB1609}&6.34&$\geq10^2$ & 5.25&0.10& 1.89&-4.44&354.6&349.3&144.7\\
1683&94&0.39&\ref{Fig:GRB1683}&1.60&2.85  & 3.59&0.08&$\leq0.02$&$<10^{-2}$&2.8&5.36&113.1\\ %%%II
1683&28&0.10&\ref{Fig:GRB1683A}&2.16&0.13 & 3.04&0.09&0.02&$\ll0.01$&25.6&36.6&328.2\\ 
1974&100&0.20&\ref{Fig:GRB1974}&1.70&$<10^{-2}$&7.31&0.53&167.0&-8.64&294.6&60.1&180.7\\
2067&29&0.08&\ref{Fig:GRB2067}&3.03&$\geq10^2$&26.3 &0.57& 17.7&-0.39& 15.7&3.6&341.7\\
2228&91&0.17&\ref{Fig:GRB2228}&18.0&$\geq10^2$&20.7&1.73&5.9&0.06&354.2&11.25&130.3\\
%\hline\hline
%\end{tabular}
%\end{flushleft}
%\end{table*}
\enddata
\label{Tab:fits}
\end{deluxetable}

\clearpage

%%\plotone{M3m4a0536ppMxw.ps}
\begin{figure}[t]\vspace{0.3cm}
\centerline{\hspace{-0.8cm}\psfig{file=M3m4a0515ppMxw.ps,height=7cm,angle=-90}}
\end{figure}

%%\plotone{Fig:Lphi}
\begin{figure}[ht]
\centerline{\psfig{file=lum_psi4.ps,height=7cm,angle=-90}}
\end{figure}

%%\plotone{Fig:Iintrinsic}
\begin{figure}[ht]
\centerline{\psfig{file=lum_int.ps,height=7cm,angle=-90}}
\end{figure}

%%plotone{Fig:0_1_02_0_10_0}
\begin{figure}[ht]
\centerline{\psfig{file=Ft0_1_02_0_10_0.ps,height=7cm,angle=-90}}
\end{figure}

%%plotone{Fig:0_1_02_2_0_0}
\begin{figure}[ht]
\centerline{\psfig{file=Ft0_1_02_2_0_0.ps,height=4cm,angle=-90}}
\centerline{\psfig{file=Ft0_1_01_2_0_0.ps,height=4cm,angle=-90}}
\centerline{\psfig{file=Ft0_1_02_4_0_0.ps,height=4cm,angle=-90}}
\centerline{\psfig{file=Ft0_4_02_2_0_0.ps,height=4cm,angle=-90}}
\centerline{\psfig{file=Ft0_4_4_12_0_0.ps,height=4cm,angle=-90}}
\end{figure}

%%plotone{Fig:0_1_02_2_12_0}
\begin{figure}[ht]
\centerline{\psfig{file=Ft0_1_02_2_12_0.ps,height=4cm,angle=-90}}
\centerline{\psfig{file=Ft0_1_01_2_12_0.ps,height=4cm,angle=-90}}
\centerline{\psfig{file=Ft0_1_02_4_12_0.ps,height=4cm,angle=-90}}
\centerline{\psfig{file=Ft0_4_02_2_12_0.ps,height=4cm,angle=-90}}
\centerline{\psfig{file=Ft0_4_4_12_12_0.ps,height=4cm,angle=-90}}
\end{figure}

%%plotone{Fig:GRB143}
\begin{figure}[ht]
\centerline{\psfig{file=GRB143E3.ps,height=9cm,angle=-90}}
\end{figure}

\begin{figure}[ht]
\centerline{\psfig{file=GRB543E3.ps,height=9cm,angle=-90}}
\end{figure}

\begin{figure}[ht]
\centerline{\psfig{file=GRB999E3.ps,height=9cm,angle=-90}}
\end{figure}

\begin{figure}[ht]
\centerline{\psfig{file=GRB1425E3.ps,height=9cm,angle=-90}}
\end{figure}

\begin{figure}[ht]
\centerline{\psfig{file=GRB1609E3.ps,height=9cm,angle=-90}}
\end{figure}

\begin{figure}[ht]
\centerline{\psfig{file=GRB1683E3II.ps,height=9cm,angle=-90}}
\end{figure}

\begin{figure}[ht]
\centerline{\psfig{file=GRB1683E3A.ps,height=4.5cm,angle=-90}}
\end{figure}

\begin{figure}[ht]
\centerline{\psfig{file=GRB1974E3.ps,height=9cm,angle=-90}}
\end{figure}

\begin{figure}[ht]
\centerline{\psfig{file=GRB2067E3.ps,height=9cm,angle=-90}}
\end{figure}

\begin{figure}[ht]
\centerline{\psfig{file=GRB2228E3.ps,height=9cm,angle=-90}}
\end{figure}

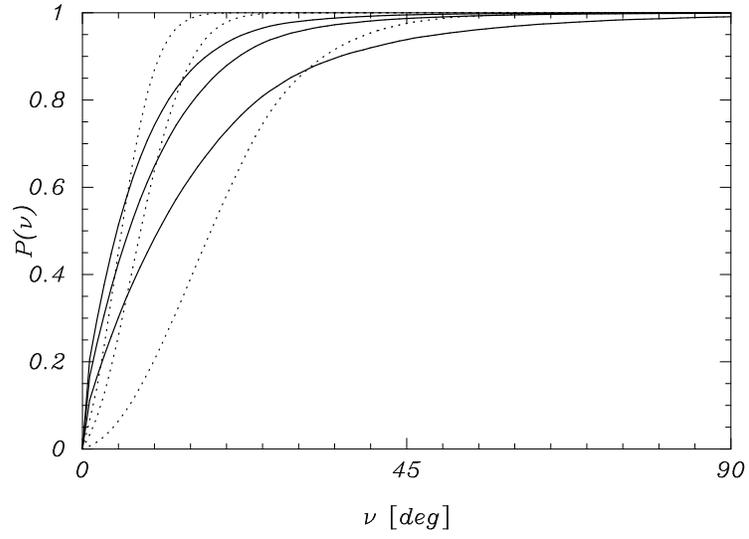
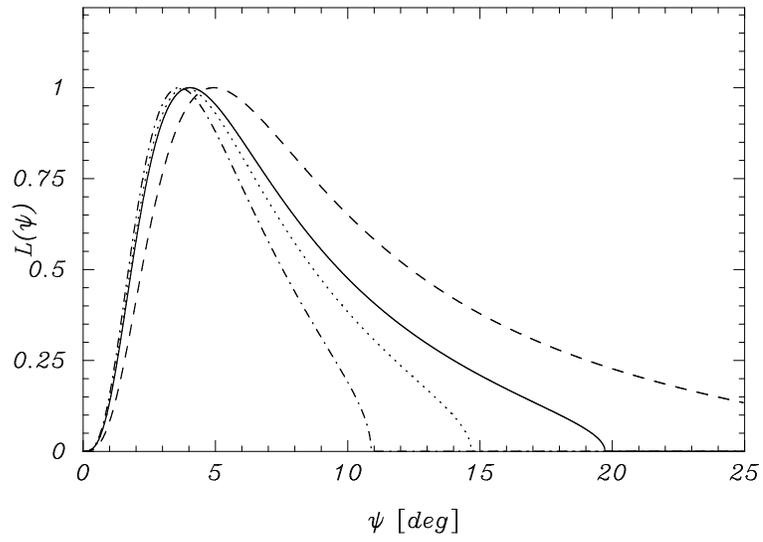
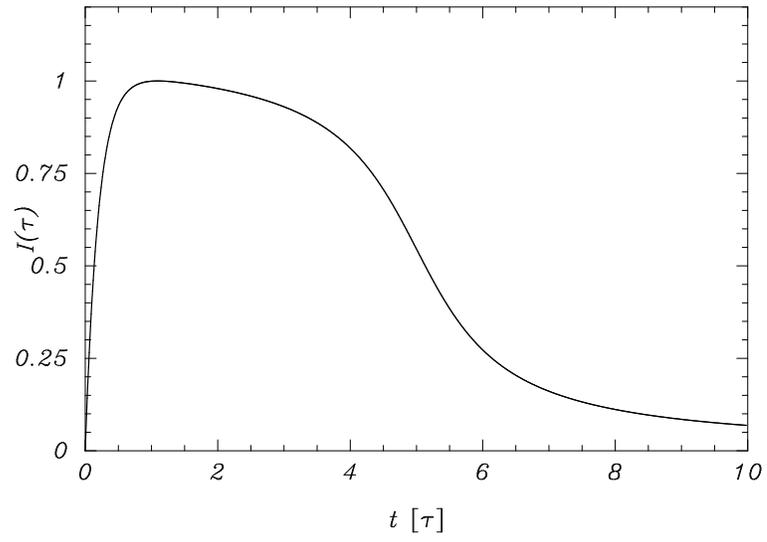
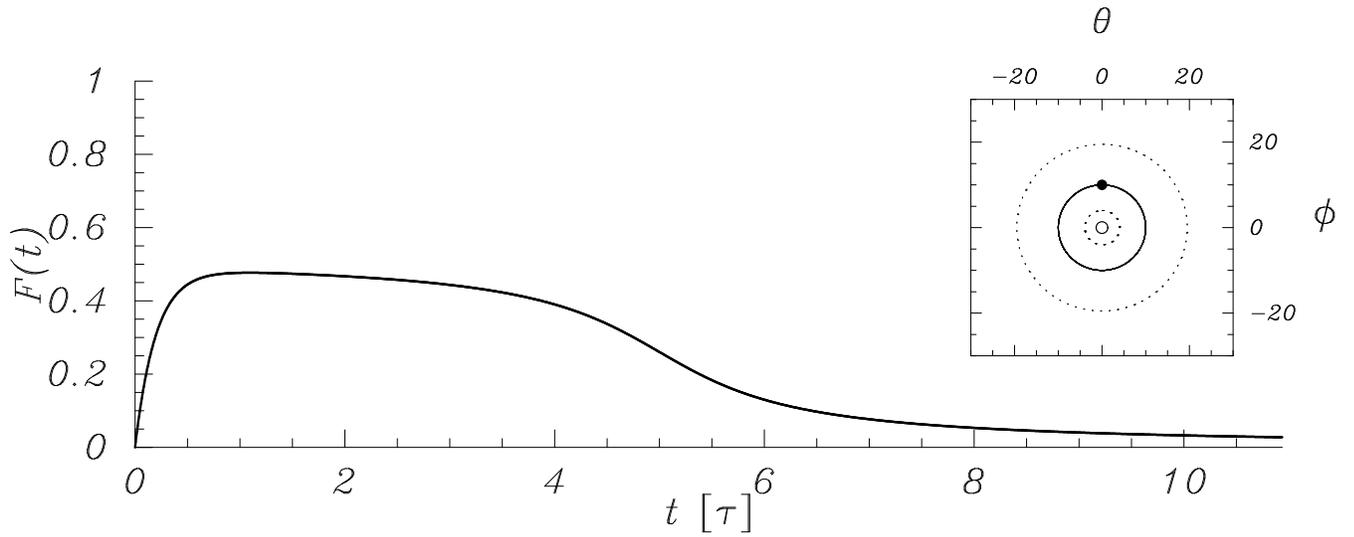
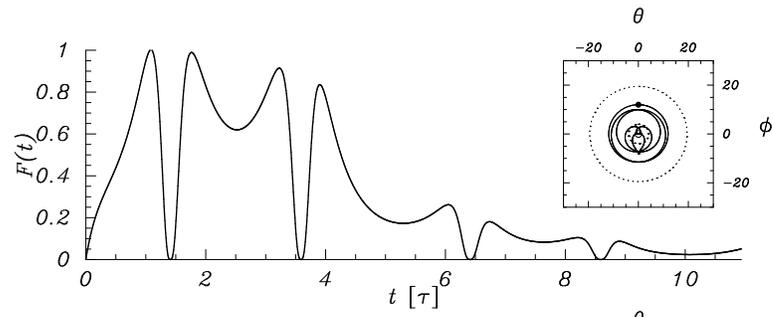
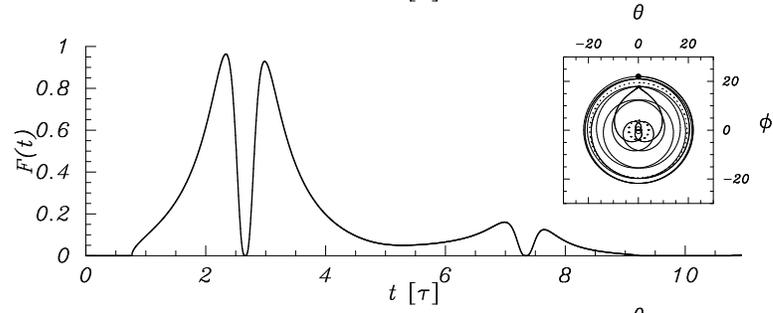
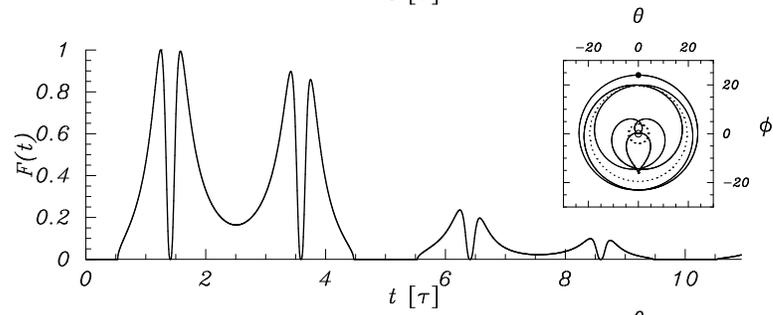
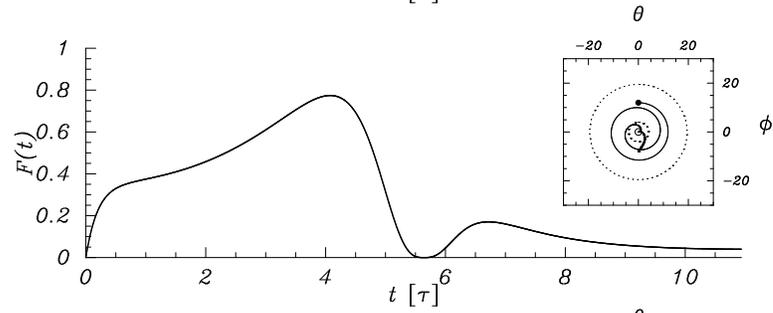
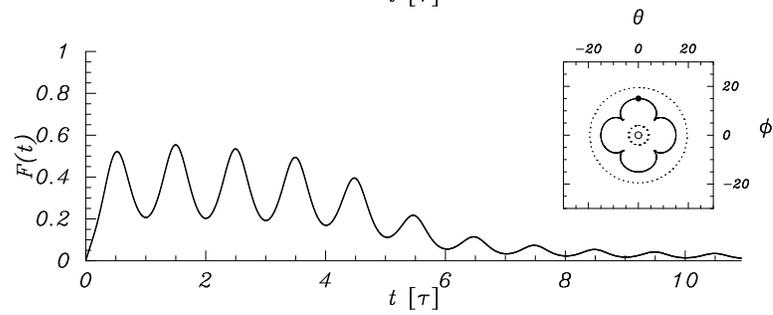
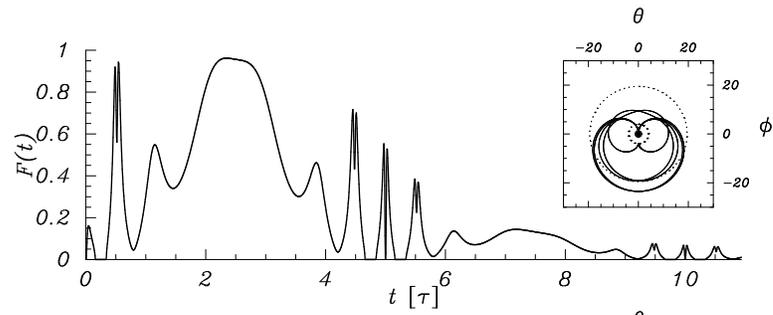
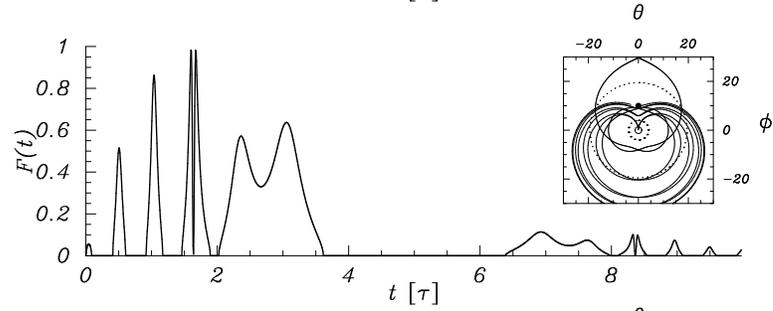
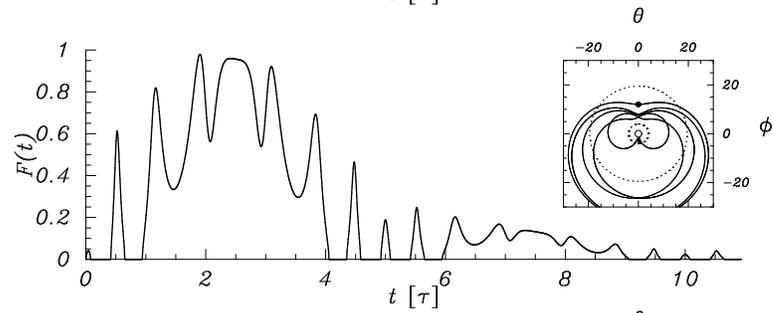
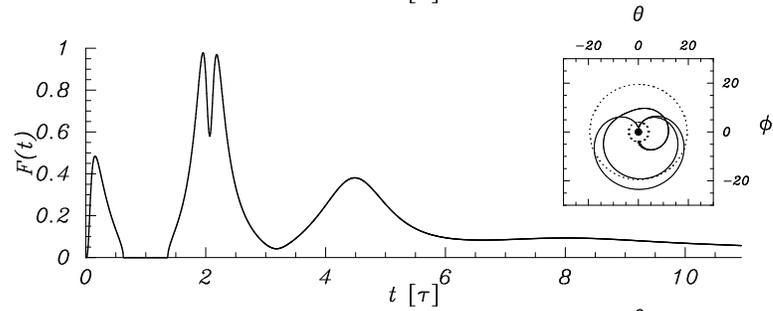
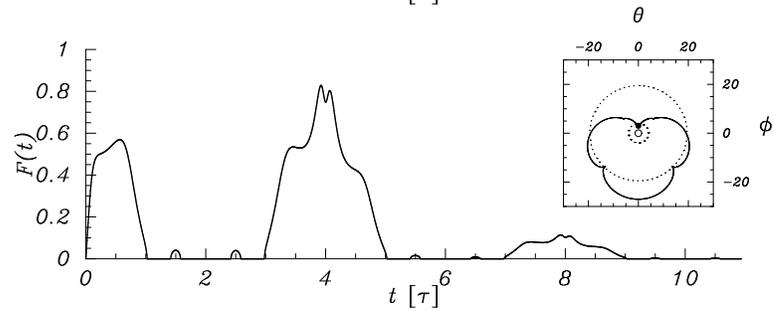
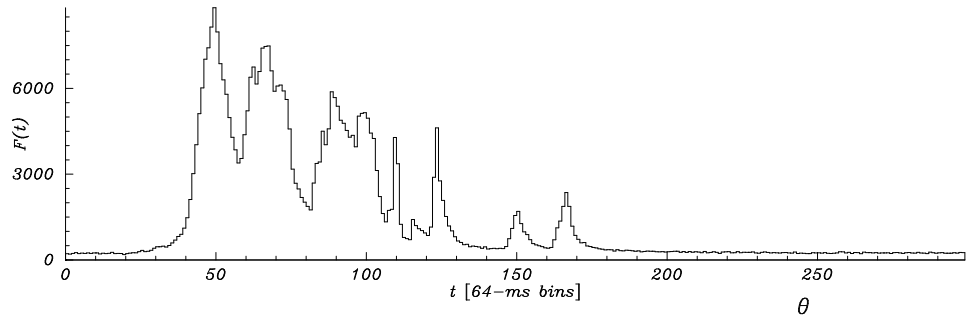
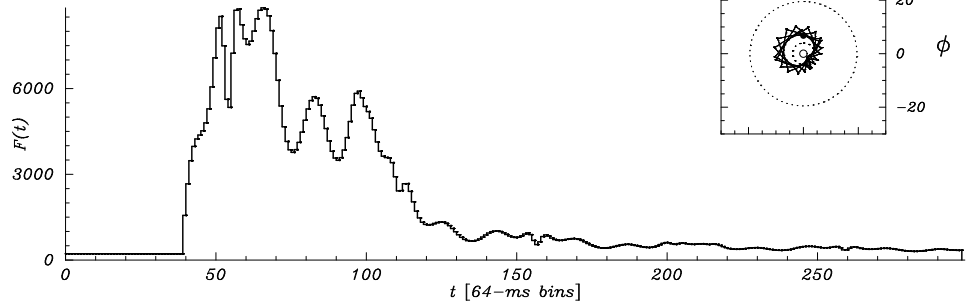
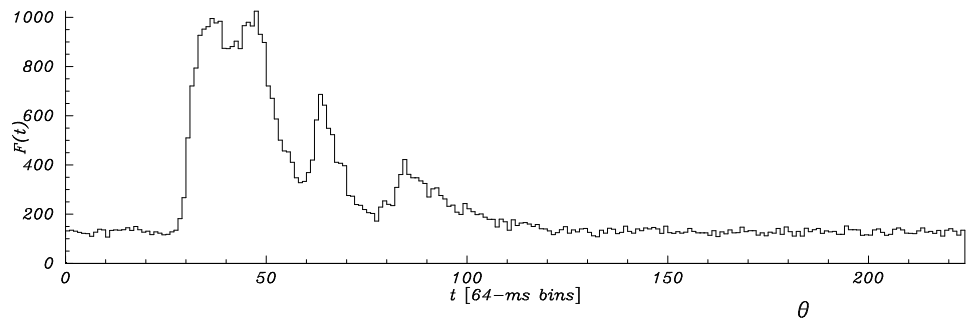
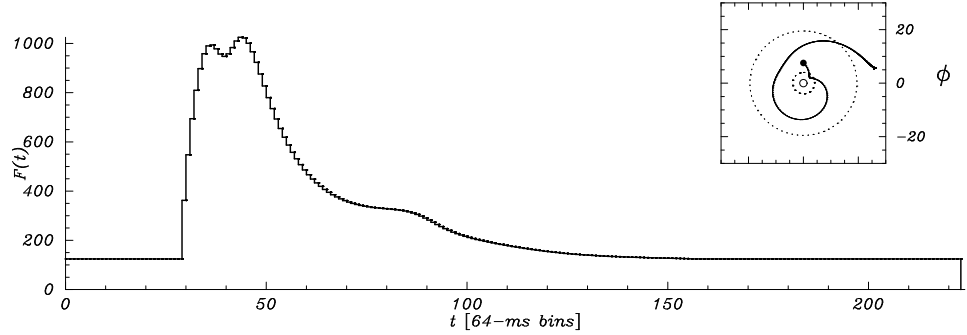
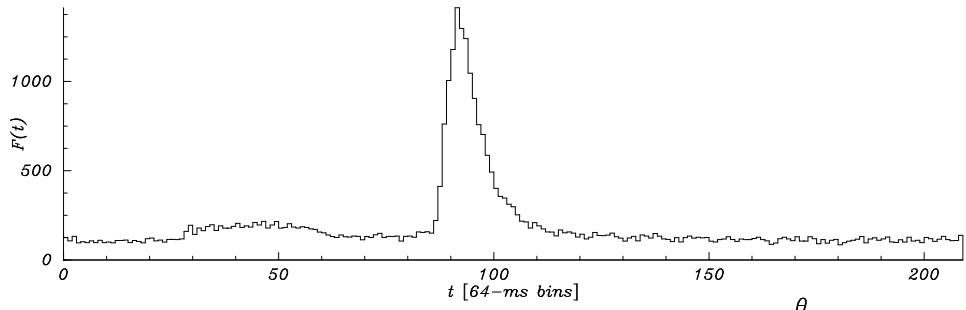
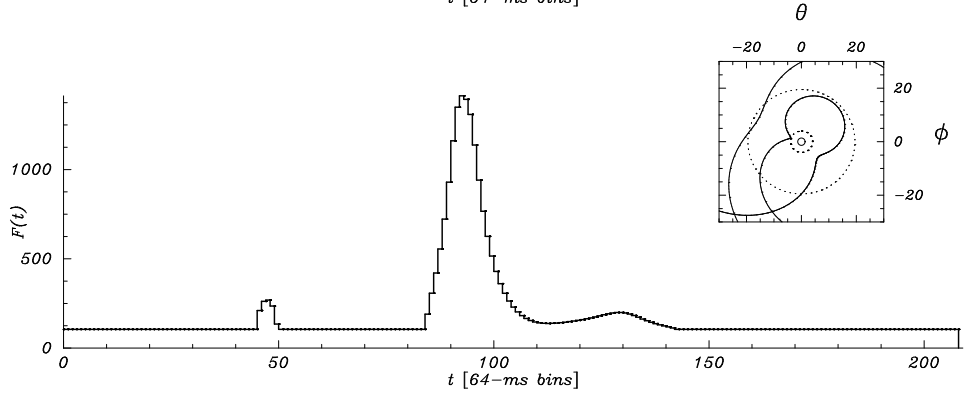
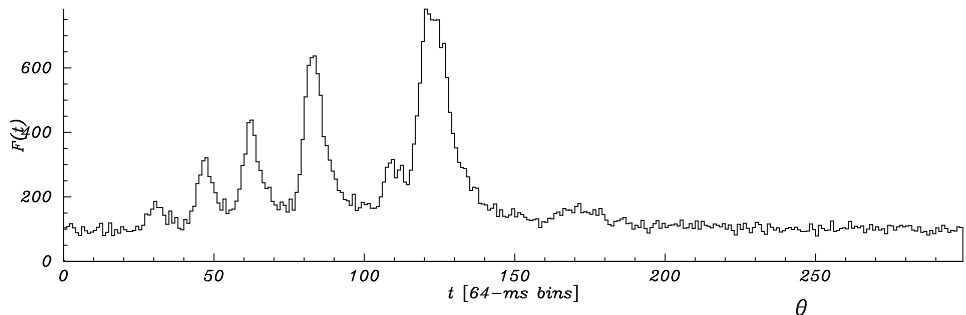
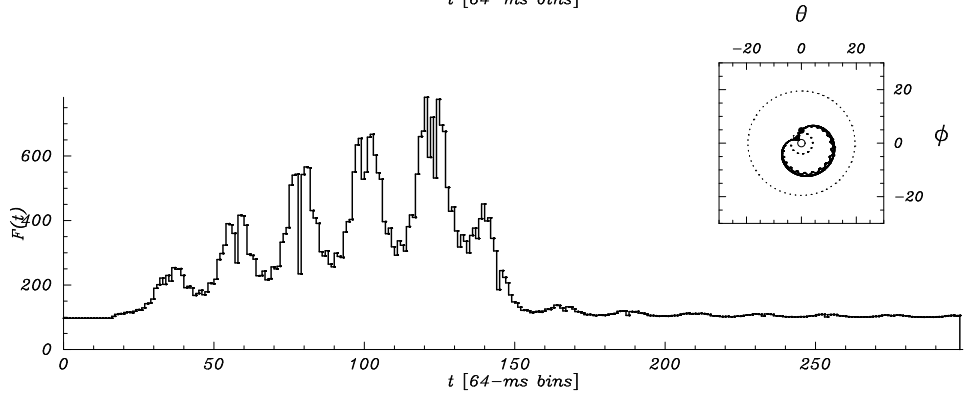
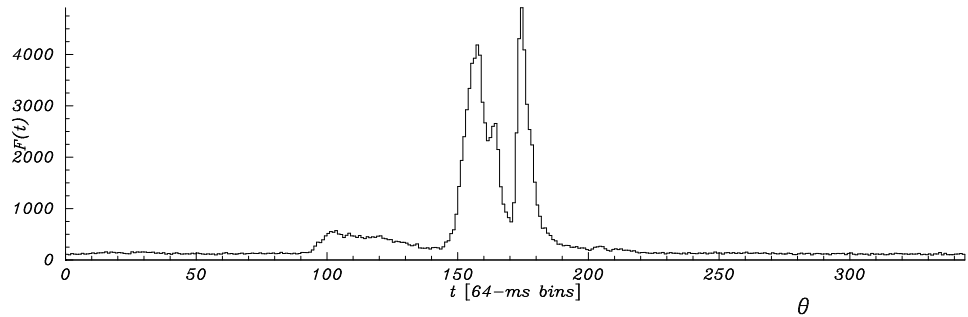
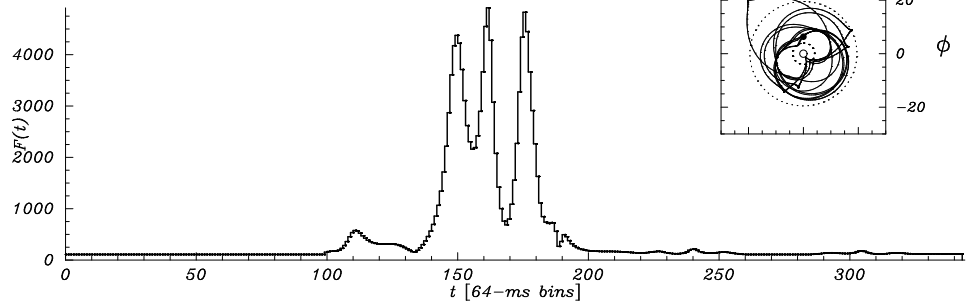
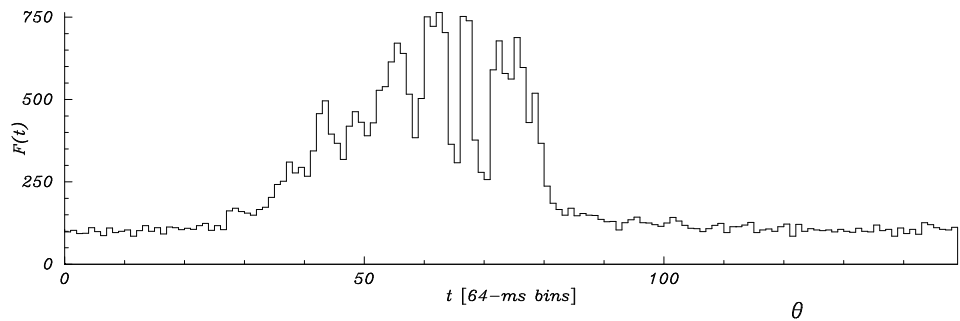
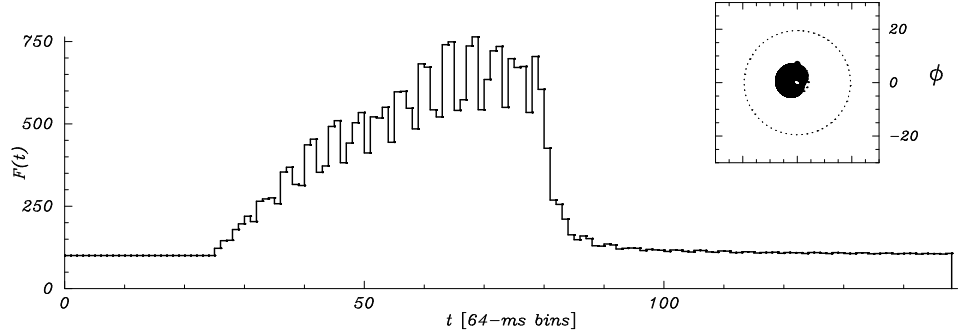
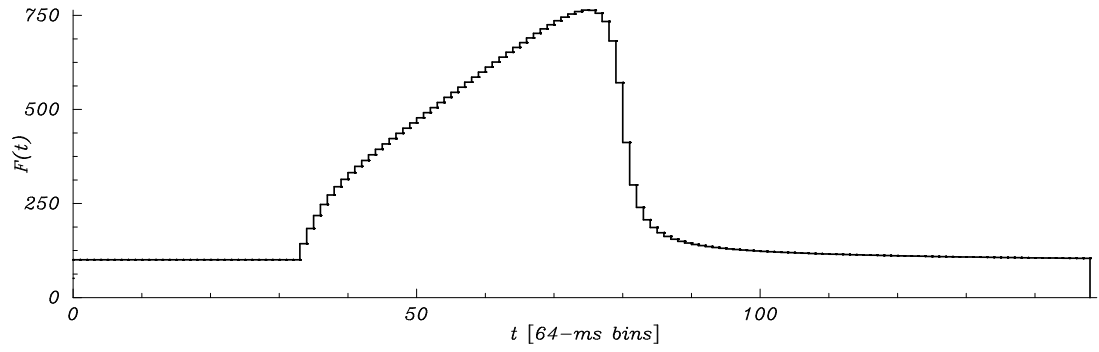
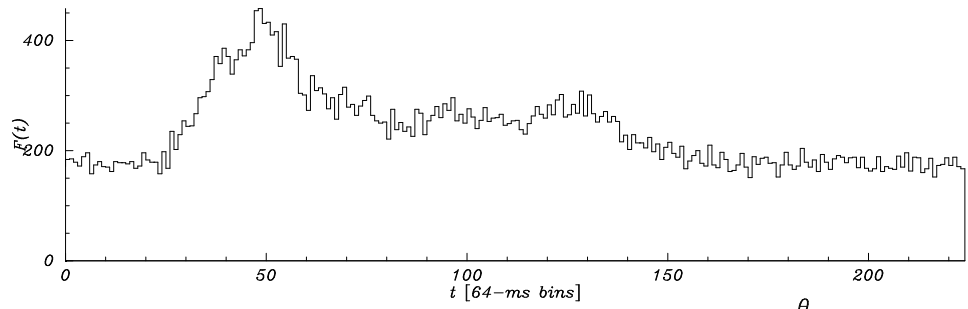
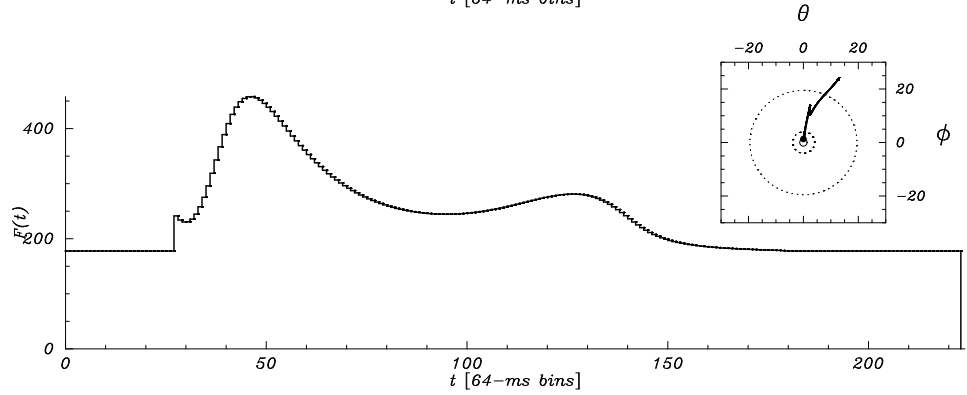
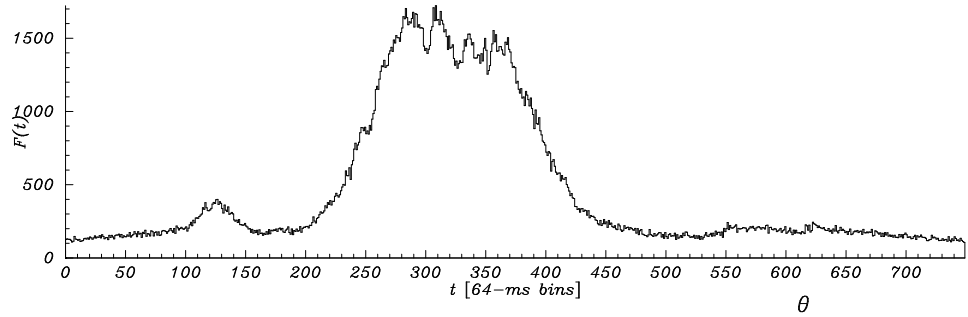
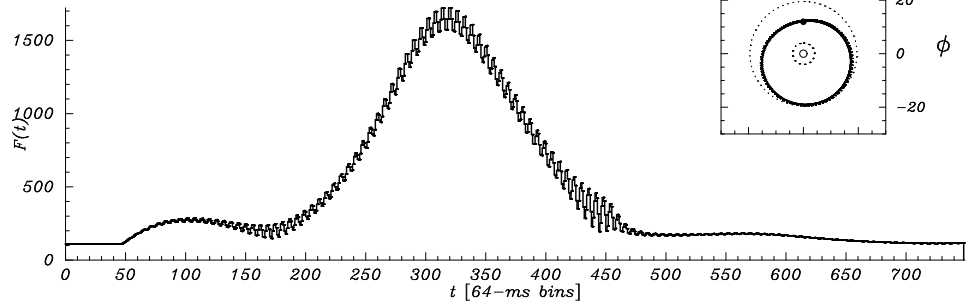
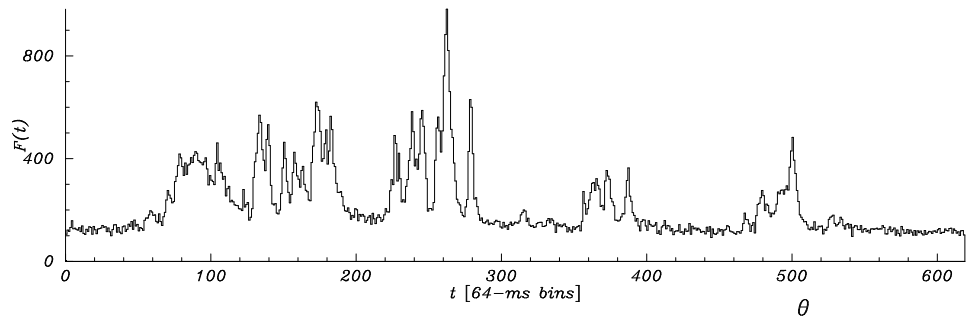
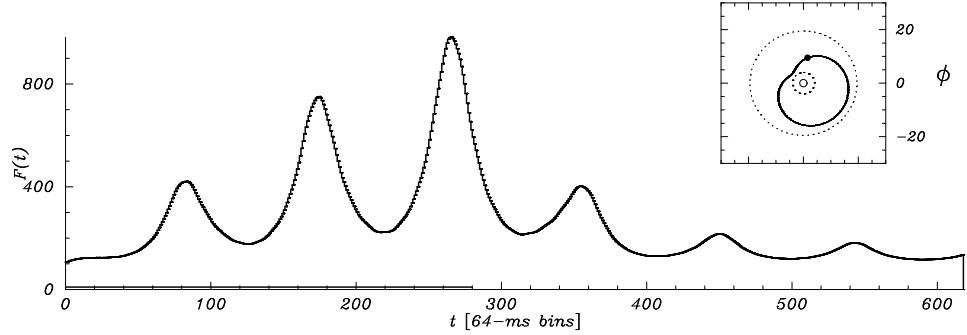
\end{document}